\documentclass[prl,aps,twocolumn,showpacs]{revtex4}
\usepackage{epsf,graphics,graphicx}
\usepackage{amsmath}
\usepackage{amssymb,latexsym,mathrsfs}
\usepackage{hyperref}
\def\bea{\begin{eqnarray}}
\def\eea{\end{eqnarray}}
\def\ba{\begin{array}}
\def\ea{\end{array}}

\def\beq{\begin{equation}}
\def\eeq{\end{equation}}
\newcommand{\nonum}{\nonumber}

\begin{document}

\title{Quantum phases and dynamics of geometric phase in a quantum spin chain
under linear quench}

\author{Sujit Sarkar}
\affiliation{PoornaPrajna Institute of Scientific Research, 4
Sadashivanagar, Bangalore 5600 80, India}
\author{B. Basu}
\affiliation{Physics and Applied Mathematics Unit, Indian
Statistical Institute, Kolkata 700 108, India}


\begin{abstract}
We study the quantum phases of anisotropic XY spin chain in presence
and absence of adiabatic quench. A connection between geometric
phase and criticality is established from the dynamical behaviour of
the geometric phase for a quench induced quantum phase transition in
a quantum spin chain. We predict XX criticality associated with a
sequence of non-contractible geometric phases.
\end{abstract}

\maketitle


{\it Introduction:} In recent times, the quantum phase transition
(QPT) has become one of the prime research topics in condensed
matter physics both from the theoretical and experimental
perspective. QPT is associated with the fundamental changes that
occur in the macroscopic nature of the matter at zero temperature
due to the variation of an external parameter. Quantum phase
transitions are characterized by the drastic change in the ground
state
properties of the system driven by the quantum fluctuations. \\
In this letter, we study the quantum phases and quantum phase
transition of exactly solvable XY spin chain model. Although the
exactly solvable XY model has been well studied \cite{barouch} and
is known to present a very rich structure but still the effect of
linear quenching on the quantum phases has not been studied
explicitly. The study of quantum phase transition and the nature of
criticality through the analysis of the dynamics of geometric phase
for linear quenching process is also rare for this model. The XY
model exhibits different regions of criticality, like XX
criticality, XY criticality and Ising criticality depending on the
values of the parameters (anisotropic exchange interaction and
magnetic field) of the system. In this letter, we study  the nature
of criticality explicitly through the dynamics of geometric phase
when the system under consideration is under quench induced QPT.  We
also address the issue on the nature of the geometric phase in the
context of XX criticality; it is shown that the XX region of
criticality is characterized by the existence
of non-contractible geometric phase.
Here we mention very briefly the essence of geometric phase in
condensed matter: Geometric phases have been associated with a
variety of condensed matter phenomena
\cite{thou,resta,hatsugai,bb1,bps,bb} since its inception
\cite{berry}. Besides, various theoretical investigations, geometric
phases have been experimentally tested in various cases, e.g. with
photons \cite{p1,p2,p3}, with neutrons \cite{n1,n2} and with atoms
\cite{a1}. The generation of a geometric phase (GP) is a witness of
a singular point in the energy spectrum that arises in all
non-trivial geometric evolutions. In this respect, the connection of
geometric phase with quantum phase transition (QPT) has been
explored very recently \cite{car,zhu,hamma}. The geometric phase can
be used as a tool to probe QPT in many body systems. Since response
times typically diverge in the vicinity of the critical point,
sweeping through the phase transition with a finite velocity leads
to a breakdown of adiabatic condition and generate interesting
dynamical (non-equilibrium ) effects. In the case of thermal phase
transitions, the Kibble-Zurek (KZ) mechanism \cite{kib,zur} explains
the formation of defects via rapid cooling. This idea of  defect
formation in second order phase transition has been extended to zero
temperature quantum phase transition (QPT) \cite{zur1,dziar} by
studying the spin models under linear quench. We will use this
concept in our study.\\
{\it{Quantum Phase Analysis and Effect of Linear Quench:}} Let us
start with the model Hamiltonian
 \beq H = \sum_{i =
-M}^{M}~(\frac{1+ \alpha }{2} {{\sigma}_i}^{x} {{\sigma}_{i+1}}^{x}
~+~\frac{1- \alpha }{2} {{\sigma}_i}^{y} {{\sigma}_{i+1}}^{y}
 + B (t) {\sigma_i}^{z} )~~~
 \eeq
 where $i$ is the site index, $x$, $y$,
and $z$ denote components of spin. $\alpha$ is the anisotropic
coupling strength and  $B$ is the
linear quench induced magnetic field in the $z$ direction.\\
We recast the spinless fermions operators
in terms of field operators by the relation
$ {\psi}(x)~=~~[e^{i k_F x} ~ {\psi}_{R}(x)~+~e^{-i k_F x} ~ {\psi}_{L}(x)] $
where ${\psi}_{R} (x)$ and ${\psi}_{L}(x) $ describe the
second-quantized fields of right- and left-moving fermions
respectively. One can express the fermionic fields in terms of
bosonic fields by the relation
$ {{\psi}_{r}} (x)~=~\frac{U_r}{\sqrt{2 \pi }}~e^{-i ~(r \phi (x)~-~
\theta (x))}, $
$r$ denoting the chirality of the fermionic fields:
 (+1) for right or (-1) for left movers.
The operators $U_r$ commute with the bosonic fields. $U_r$ of
different species commute and $U_r$ of the same species anticommute.
$\phi$ field corresponds to the quantum fluctuations (bosonic) of
spin and $\theta$ is the dual field of $\phi$. They are related by
the relation $ {\phi}_{R}~=~~ \theta ~-~ \phi$ and  $ {\phi}_{L}~=~~
\theta ~+~ \phi$. We finally get the bosonized Hamiltonian as \bea H
&=& {H_0} ~+~ \frac{{\alpha}}{2 {\pi}^2 {a}^2}\int dx~
cos[2 \sqrt{\frac{\pi}{K}} \theta (x)] \nonum\\
& & + B(t) \sqrt{\frac{\pi}{K}} \int dx~ {{\partial}_x} \phi (x).
\eea In this derivation, we have used the following expressions for
spin operators in terms of the bosonic fields:
$ S_n^x  =  [~ c_2 \cos (2 {\sqrt {\pi K}} \phi) ~+~ (-1)^n c_3 ~]~
\cos ({\sqrt {\frac{\pi}{K}}} \theta )$ , $ S_n^y  =  -[~ c_2 \cos
(2 {\sqrt {\pi K}} \phi) ~+~ (-1)^n c_3 ~]~
 \sin ({\sqrt {\frac{\pi}{K}}} \theta )$,
$ S_n^z  =  {\sqrt {\frac{\pi}{K}}} ~\partial_x \phi ~+~ (-1)^n c_1
\cos (2 {\sqrt {\pi K}} \phi )$ . $K$ is the Luttinger liquid (LL)
parameter,
$c_2$ and $c_3$ are the constants.\\
Now we analyse the quantum phases of the system in the absence of linear
quench:
The second term of the Hamiltonian, which is the sine-Gordon
coupling term  is relevant when $K >1/2 $. The elementary excitation
is gapped (one can estimate the mass gap by this relation $ M =
\Lambda {(\frac{\alpha}{2})}^{\frac{1}{2 - 1/K}} $ , $\Lambda$ is a
cutoff parameter) and the system is in the staggered order phase. On
the otherhand, for $K < 1/2 $,
the system is in the Luttinger liquid phase.\\
Next we interpret the quantum phases of the system by analyzing the
two renormalization group equations.  The renormalization group (RG)
equation for the Hamiltonian (Eq. 2) is \beq \frac{d \alpha}{dl}  =
 (2 - 1/K) \alpha,~ \frac{dK}{dl}  =   {{\alpha}^2}/4  . \eeq
We interpret from the first RG equation that sine-Gordon coupling
with strength $\alpha$ will be relevant and the system will flow
from the weak coupling phase to the strong coupling phase, when $K>
1/2 $ otherwise this coupling term is irrelevant and the system is
in the LL phase of the system. The second RG equation reveals that
as $\alpha $ will increase, the LL parameter will increase, i.e, the
flow of the second RG equation also support the flow of the first RG
equation for its relevant
phase, i.e., the system is in the staggered order phase.
Results derived from Abelian bosonization method is consistent
with the RG study because these RG equations have only trivial
fixed point.\\
Now we analyze the effect of linear quenching (last term of Eq. 1) ,
it modifies the quantum phases of the system. In the static limit
for $ K < 1/2$, the system is in the LL phase. The system drives to
the ferromagnetic phase due to the presence of linear quench
induced magnetic field in the z direction.\\
In the static limit, sine-Gordon coupling term is relevant ($ K> 1/2
$), the system has a excitation gap (stagger phase). The system
drives to the ferromagnetic phase when the quench induced magnetic
field is larger than the excitation gap of the system. The system is
in the LL phase when the quench induce magnetic field is of the
order of magnitude of the gap ($ B(t) \sim  M = \Lambda
{(\frac{\alpha}{2})}^{\frac{1}{2 - 1/K}} $). At time $t =0$,
system is in the staggered order magnetic phase.\\
{\it Geometric Phase and Criticality}:
In this model, the geometric phase of the ground state is evaluated
by applying a rotation of $\phi$ around the
$Z$-axis in a closed circuit to each spin \cite{car, car1}.
A new set of Hamiltonians $H_\phi$ is constructed from the Hamiltonian
 (1) as
 \begin{equation}\label{h2}
 H_\phi=U(\phi)~H~U^\dagger(\phi)
 \end{equation}
 where
$U(\phi)=\prod_{j=-M}^{+M} \exp(i\phi\sigma_j^z/2) $
 and $\sigma_j^z$ is the $z$ component of the standard Pauli matrix at site $j$.
 The family of Hamiltonians generated by varying $\phi$ has the same energy spectrum as
the initial Hamiltonian and $H(\phi)$ is $\pi$-periodic in $\phi$.
 With the help of standard Jordan-Wigner transformations,
which makes the spins to one dimensional spinless fermions via the
relation $a_j=\left( \prod_{i<j}\sigma_i^z \right) \sigma_j^\dagger$
 and then using the Fourier transforms of the fermionic operator,
$d_k=\frac{1}{\sqrt{N}}\sum_j ~a_j\exp\left( \frac{-2\pi j_k}
{N}\right)~~~{\rm{with}}~~~ k=-M,...+M$
the Hamiltonian $H_\phi$ can be diagonalized by transforming the fermionic
operators in momentum space and then using
 Bogoliubov transformation.
The ground state  $|g> $  of the system is expressed as
 \begin{equation}\label{g}
 |g> =\prod_{k>0}(\cos\frac{\theta_k}{2}|0>_k |0>_{-k}
-ie^{2i\phi}\sin\frac{\theta_k}{2}|1>_k|1>_{-k} )
 \end{equation}
 where $|0>_k$ and $|1>_k$ are the vacuum and single fermionic excitation of the
$k$-th momentum mode respectively. The angle $\theta_k$ is given by
\begin{equation}
\cos\theta_k=\frac{\cos k-B}{\Lambda_k}
 \end{equation}
 and
$ \displaystyle{ \Lambda_k=\sqrt{(\cos k-B)^2+\alpha^2\sin^2 k}} $
 is the energy gap above the ground state.
 The ground state is a direct product of $N$ spins, each lying in the
two-dimensional Hilbert space spanned by $|0>_k
 |0>_{-k}$ and $|1>_k |1>_{-k}$. For each value of $k$,
the state in each of the two dimensional Hilbert space can be represented
as a Bloch vector with coordinates $(2\phi,\theta_k)$.
The overall phase is given by the sum of the individual phases.
One can also write the above Hamiltonian as single particle excitations
$H (\alpha, B(t), \phi) = \sum_{-M}^{M} {\Lambda}_k {b_k}^{\dagger}
{b_k} $, where
$b_k = cos({{\theta}_k}/2) d_k -
i e^{2i \phi} sin({{\theta}_k}/2 ) {d_k}^{\dagger} $.
The pseudomomenta $k$ take $half$ $integer$ values:
$ k=\pm\frac{1}{2}\frac{2\pi}{N},.....,\pm\frac{N-1}{2}\frac{2\pi}{N}.$
The direct calculation shows that the geometric phase for the $kth$  mode,
which represents the area in the parameter space enclosed by the loop
determined by ($2\phi,\theta_k)$ is given by
 \begin{equation}\label{ph}
  \Gamma_k=\pi(1-\cos\theta_k)
\end{equation}
 The geometric phase of the state $|g>$ is given by
$ \Gamma_g=\sum_{k}~ \Gamma_k. $
For an adiabatic evolution, if the initial state is an eigenstate,
the evolved state remains in the eigenstate.  So 
 we may now derive the instantaneous geometric phases of this system
due to a gradually decreasing magnetic field.
Let us  explore the situation when  the system (1) is driven adiabatically
(slow transition ) by a  time dependent magnetic field $B(t)$ such that
\begin{equation}\label{quench}
 B(t<0)=-\frac{t}{\tau_q}
 \end{equation}
 $B(t)$, driving the transition, is assumed to be linear with an
adjustable time parameter $\tau_q$ ($1/{\tau_q} $ is the quenching
rate).
 Let the system be initially at time $ t(<0)<<\tau_q$ such that $B(t)>>1$.
 The instantaneous ground state at any instant $t$
is given by
\begin{equation}\label{g1}
 |\psi_0(t)>=\prod_{k}(\cos\frac{\theta_k(t)}{2}|0>_k|0>_{-k}
-ie^{(2i\phi)}\sin\frac{\theta_k(t)}{2}|1>_k|1>_{-k})
 \end{equation}
We now use eqn. (7) and (6) to derive
 the geometric phase of the $k^{th}$ mode  which yields
 \begin{equation}
\Gamma_k(t)=\pi\left(1-\frac{\cos k +\frac{t}{\tau_q}}
{\sqrt{(\cos k +\frac{t}{\tau_q})^2+\alpha^2\sin^2 k}}\right)
\end{equation}
The geometric phase for an isotropic system with $\alpha=0$ and
quantum Ising model with $\alpha=1$ may now be easily obtained.
\begin{eqnarray}
 \rm{For}~ \alpha  =  0, ~\Gamma_k(t)& = & 2 \pi \Theta(|t| -{{\tau}_q})~~  {\rm{and}}\nonumber \\
  \rm{for}~ \alpha  =  1,~
 \Gamma_k(t)  &= &\pi \left( 1-\frac{\cos k
 +\frac{t}{\tau_q}}{\sqrt{1+\frac{t^2}{\tau_q^2}+2\frac{t}{\tau_q}\cos k}}\right)
\end{eqnarray}
For a system of size $N$ the total geometric phase for the initial state is,
$ \Gamma_{initial}=\sum_k \Gamma_{k}. $
The magnetic field is gradually decreased by adjusting $\tau_q$ and the critical point is
attained at $t=-{\tau_q}$ with $B=1$. Then the geometric phase for the k th mode is
\begin{equation}
 \Gamma_k(t=-\tau_q) = \pi \left(1-\frac{\cos k+1}{\sqrt{(\cos k +1)^2+\alpha^2\sin^2 k}}\right)
 \end{equation}
 At the critical point the total geometric phase is
$ \Gamma_{critical} = \sum_k \Gamma_{k}(t= -{\tau_q} ). $
Finally, at $t=0$, when the magnetic field is gradually turned off,
the situation is a bit different. The configuration of the final
state will depend on the number of kinks generated in the system due
to phase transition at or near  $t=-{\tau_q}$ and as such it will
depend on the quench time $\tau_q$ \cite{zur}. The number of kinks
is the number of quasi-particles excited at $B=0$ and is given by
$ {\cal{N}}={\sum_k }p_k $
where $p_k$, the excitation probability (for the slow transition)
is given by the Landau Zener formula \cite{zener}
$  p_k \approx \exp{ (-2\pi \tau_q k^2)} $.
As different pairs of quasi-particles ($k,-k$) evolve independently,
for large values of $\tau_q$, it is likely that only one pair of quasi-particles
with momenta $(k_0,-k_0)$ will be excited where $k_0(=\frac{\pi}{N})$ corresponds
to the minimum value of the energy $\Lambda_k$.
Thus the condition for adiabatic transition in a finite chain is given by,
$ \tau_q >> \frac{N^2}{2 \pi^3} $.
Hence, well in the adiabatic regime, the
final state at $t=0$  is given by
\begin{eqnarray*}
 |\psi_{final}>  = |1>_{k_0}|0>_{-k_0}\\
  \prod_{k,k\neq\pm k_0}
 (\cos\frac{\theta_k}{2}|0>_k|0>_{-k}
 -ie^{(2i\phi)}\sin\frac{\theta_k}{2}|1>_k|1>_{-k})
 \end{eqnarray*}
 This state is similar to direct product of only $N-1$ spins oriented along $(2\phi, \theta_k)$
where the state of the spin corresponding to momentum $k_0$ does not contribute to the geometric phase.
The total geometric phase of this state is given by
$ \Gamma_{final}(t=0)=\sum_{k, k\neq \pm k_0}\pi (1-\cos\theta_k) $
\begin{figure}
\includegraphics[scale=0.40,angle=0]{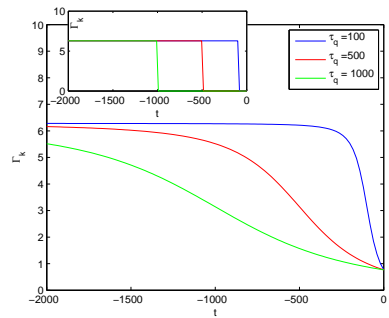}
\caption{Color online, variation of geometric phase with time for
different quenching time. The anisotropy parameter ($\alpha =0.5$).
Inset shows the same study but for $\alpha =0$ (XX model).
}
\label{Fig. 1 }
\end{figure}
\begin{figure}
\includegraphics[scale=0.50,angle=0]{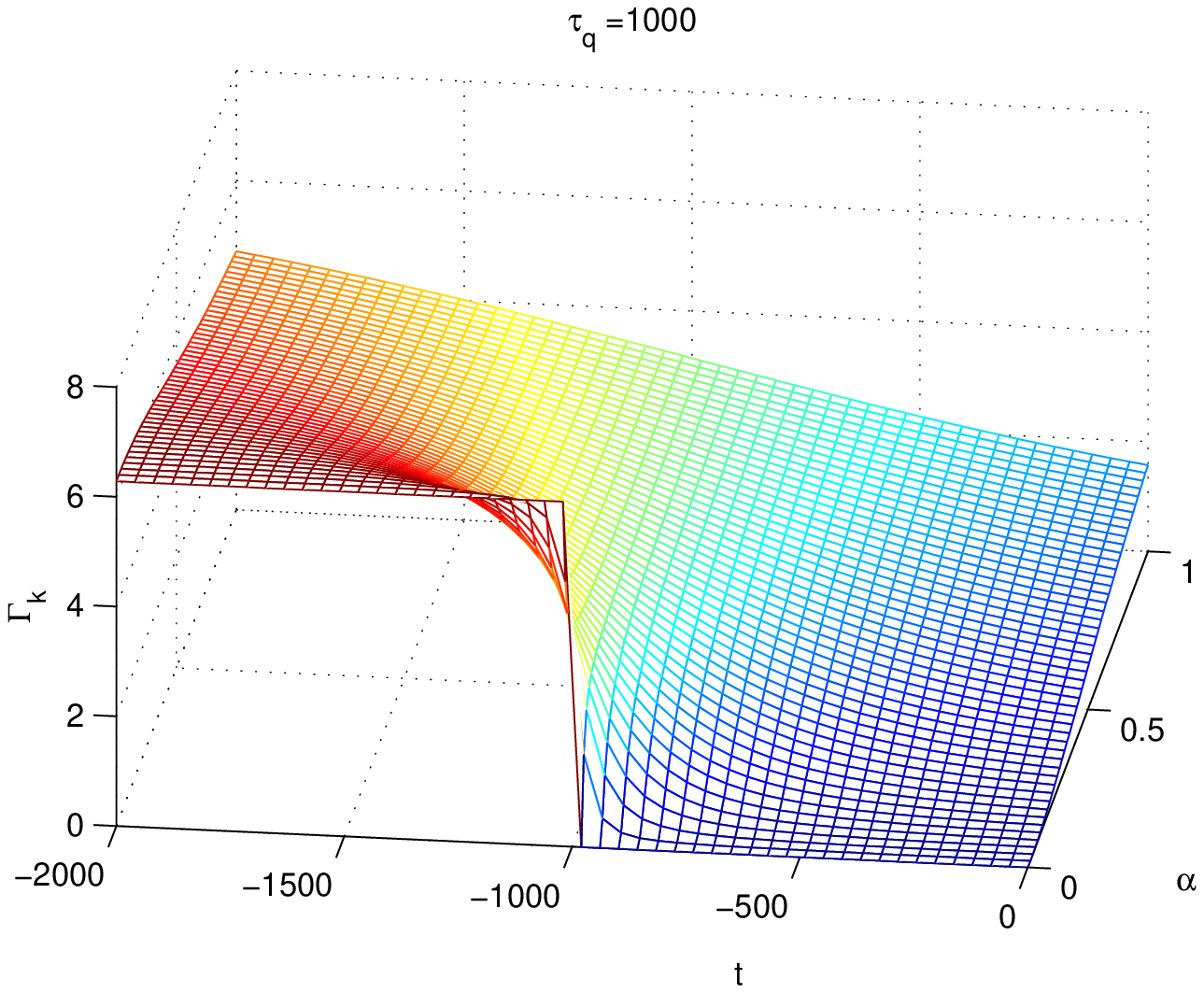}
\includegraphics[scale=0.50,angle=0]{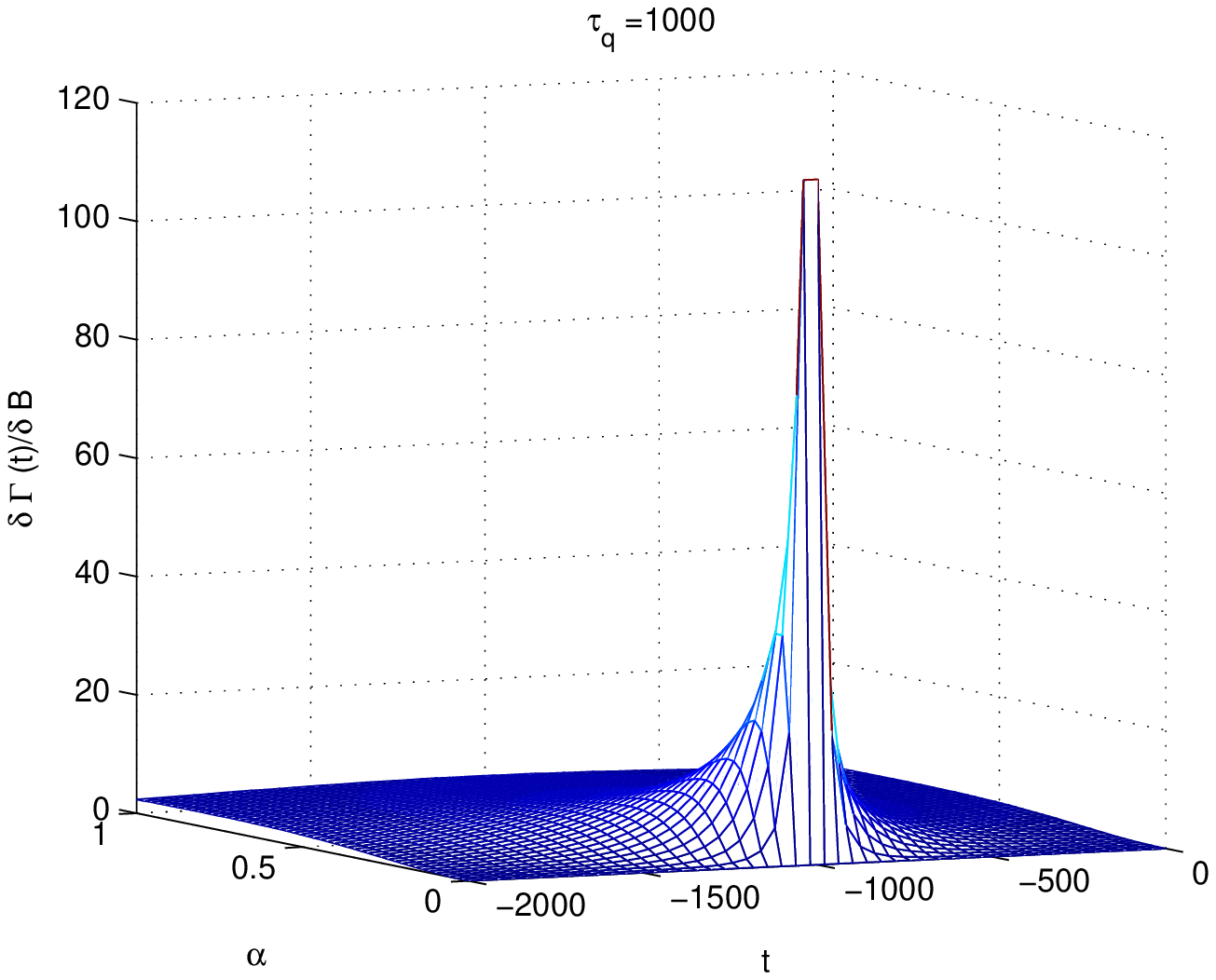}
\caption{Color online, the left planel of this figure is for the
geometric phase and right one is for the derivative of the geometric
phase to study the quantum criticality. We present the vartion of
these quantities with anisotropy parameter and time.
 }
\label{Fig. 2 }
\end{figure}
Now we study the geometric phase associated with the quench induced
quantum phase transition.  We study the variation of geometric phase
(${\Gamma}_k $) and its derivative with respect to the quench
induced magnetic filed ($B$) i.e. $ ({\frac{d {\Gamma}_k}{d B}})$
with time. We find the non-analytic behavior of the derivative at $
t={\tau}_q$. The analytical expression for the derivative is \beq
{\frac{d {\Gamma}_k (t)}{d B}} = \frac{\pi {\alpha}^2 sin^{2} (k) }
{{({( {cos(k) + t/{{\tau}_q})}^2 + {\alpha}^2 sin^{2} (k) })}^{3/2}}
\eeq  In Fig. 1 we study the Berry phase $\Gamma_k$ for different
quench time $\tau_q$,  for the XX and XY spin chain system. For the
XX (inset) model, we find the step function like behaviour at $t =
{\tau_q} $ where the sharp transition occurs. It can be seen that
the system shows quantum criticality at that point. The sharp
transition disapperas for the finite anisotropy (XY) model. We
observe from our study that the variation
of $\Gamma_k$ become flat for slower quenching rate.\\
The upper panel of Fig. 2 shows the variation of $\Gamma_k $  with
$\alpha$ and $t$ to get the whole picture of variation of $\Gamma_k.
$ It reveals from our study that sharp transition occurs for $\alpha
=0$ only. The lower panel of the Fig. 2, shows the total variation
of $ {\frac{d {\Gamma}_k (t)}{d B}}$. The non-analytical behaviour
for $\alpha =0$ at $t= -{\tau_q} $ helps us to predict XX
criticality under the linear quenching process. The analyis with
different values of $\tau_q$ shows that the appearance of XX
criticality is independent of $\tau_q $, i.e.,
independent of fast and slow quenching rate.\\
Finaly,  we discuss  an important aspect of the XX criticality. It
is shown that the XX region of criticality is characterized by the
existence of non-contractible geometric phase. The system is in the
gapless excitations in the XX region of criticality (the gapless
excitation is defined as ${b_k}^{\dagger}$ operator). Here
$cos(\frac{2 \pi k_0}{N}) = B(t) (\Lambda_{k_0} )$, $k_0 \epsilon
{1,...M}$. For finite $M$, we can write $ \lim_{\alpha \rightarrow
0} cos(\theta_k) = \pm 1 $ and ground state
$|g(\alpha,B(t),\phi)> = \otimes_{k< k_o} |0>_k |0>_{-k}
\otimes_{k> k_0} |1>_k |1>_{-k} $. The ground state of the system
acquires the geometric phase due to the variation of $\phi$ for
fixed $\alpha$ and $B(t)$. $|g> $ is the tensor product of $M$
qubit states. We observe from the Fig. 1 and also from the
analytical expression for ${\Gamma}_k$ that $\lim_{\alpha
\rightarrow 0} {\Gamma}_k= 0, 2 \pi $. In the thermodynamic limit
there is a solution, $cos \theta_{k_0} = 0 $, which leads to the
result ${\Gamma}_k = \pi$ for every $\alpha > 0$ and hence
$\lim_{\alpha \rightarrow 0} {\Gamma}_{k_0} = \pi$. Therefore by
using the relation of total geometric phase, we can write
${\lim_{\alpha \rightarrow 0}} {\lim_{M \rightarrow \infty}}
\frac{{\Gamma}_g}{M} \neq 0$ which shows  the sequence
${{{\Gamma}_k}({\alpha}_n)}_{n \epsilon N}$ is non-contractible in
the thermodynamic limit. Hence, the non-contractible nature of the
geometric phase associated with the $XX$ criticality is proved.

To conclude, we have studied the various quantum phases of the $XY$
spin model and also  the effect of linear quench on these quantum
phases. The dynamics of the associated geometric phases are studied
for slow and rapid quenching and an intimate connection between
geometric phase and quantum criticality is  established. We have
predicted XX criticality
and also non-contractible geometric phase at criticality.\\
\\
The author (SS) would like to acknowledge The Center for Condensed
Matter Theory of IISc  and also the Physics and Applied Mathematics
Unit of I.S.I. Kolkata for providing extended facility.

\end{document}